\documentclass[12pt]{article}
\usepackage{amsmath, amsthm, amssymb,bbm}

\title{\bf Quantum measuring processes for trapped ultracold bosonic gases}

\author{S. Anderloni$^{a,b}$, F. Benatti$^{a,b}$, 
R. Floreanini$^{b}$, A. Trombettoni$^{b,c}$\\
\small ${}^a$Dipartimento di Fisica Teorica, Universit\`a di Trieste, 
34014 Trieste, Italy\\
\small ${}^b$Istituto Nazionale di Fisica Nucleare, Sezione di Trieste,
34014 Trieste, Italy\\
\small ${}^c$Scuola Internazionale Superiore di Studi Avanzati, 34014 Trieste, Italy}

\date{\null}

\begin{document}

\maketitle

\begin{abstract}
\noindent
The standard experimental techniques usually adopted in the study
of the behaviour of ultracold atoms in optical lattices involve extracting the
atom density profile from absorption images of the atomic sample after trap release.
Quantum mechanically this procedure is described by a generalized measure (POVM);
interference patterns found in absorption images suggest a generalized measure
based on fixed-phase, coherent-like states. We show that this 
leads to an average atomic density which differs from the usually adopted one,
obtained as the expectation value of the atom density operator in the many-body state.
\end{abstract}

\vskip 1 cm

\section{Introduction}

Experiments with ultracold atoms trapped in optical lattices have reached high degrees
of accuracy and sophistication in exploring the coherence effects that such systems exhibit;
they turn out to be a unique laboratory for experimental tests of many effects
in many-body quantum physics and in particular in the study of quantum phase transitions
(for recent reviews see \cite{Lewenstein,Bloch}, and references therein).

In all these studies, the technique used to extract information on the system is essentially the same
and is based on the analysis of interference phenomena (see \cite{Stringari}-\cite{Polkovnikov}
and references therein). 
Being rather difficult to measure relevant system observables directly inside the optical lattice, 
a commonly adopted experimental procedure consists in releasing the confining trap
and subsequently obtaining absorption images of the expanding atomic clouds. 
Since during expansion the atoms are usually 
considered non interacting and thus following a free, ballistic evolution, the set of the obtained
absorption images contain information on the dynamics of the atoms in the lattice prior to the
trap release. The imaging process is destructive: the sample is lost after
the measurement. Then, one usually repeats the whole procedure, preparing the system
in the same initial state, and obtaining at the end a set of absorption images.
When all images are put together and superimposed, in the limit of a large
number of them, they are usually assumed to reproduce the atom density obtained by
averaging the associated operator in the given many body system state.

In general, the quantum mechanical description of the averaging procedure
corresponds to a generalized measure, 
a so-called Positive Operator Valued Measure (POVM) \cite{Takesaki,Peres,Alicki}.
In each single image obtained by ``taking a photograph'' of the expanding clouds,
the process of photon detection gives rise to interference patterns showing a phase difference 
among different wells, even though the initial state does not have a definite phase difference: 
the imaging process effectively projects the state of the system 
into a coherent-like state, with definite phase relations, although varying randomly
from picture to picture \cite{Leggett,Stringari}. 
Given this, one can compute using the corresponding POVM the atom density profile: 
we show it to be in general different from 
the one obtained through a direct average of the operator density.

In most cases this difference turns out to be negligible due to the large number of atoms involved.
Nevertheless, at least in principle, it can be made apparent by preparing the system
in suitable states, for which the number of atoms in the various lattice sites is highly unbalanced.
In this respect, a very promising instance in which this difference might be experimentally
investigated is provided by the measure of density-density correlations in two-color
optical lattices \cite{Fort1,Fort2}.

In the following, we shall discuss the case of an optical lattices formed by just two sites,
{\it i.e.} a system of ultracold atoms confined in a double-well potential: besides simplifying
the discussion, this case has direct experimental relevance; generalization of our considerations to
a generic lattice presents no difficulties.

We shall first briefly recall the two-mode description of the trapped system and then analyze
its behaviour after the release of the trapping potential using a second-quantized
many-body approach. The average density profile obtained by superimposing the various absorption
images will then be computed starting from an initial Fock state, containing a definite
number of atoms in each well, assuming, as often done in the literature,
a free expansion of the atomic cloud after trap release. 
We shall see that for a very unbalanced initial state,
this average differs in an experimentally testable way from the standard density mean value.

\section{Cold bosonic gas in a double well trap}

In a suitable approximation, {\it i.e.} for large enough barrier,
the dynamics of cold atoms confined in a double well potential
can be described by a two-mode Bose-Hubbard type Hamiltonian \cite{Javanainen2,Jaksch,Milburn}
\begin{equation}
\label{BH}
H=E_1\, a^\dag_1a_1\,+E_2\, a^\dag_2a_2\,+\,
U\,\Bigl[(a^\dag_1)^2\, a_1^2+(a^\dag_2)^2\, a_2^2\Bigr]\,-\,T\,(a^\dag_1a_2+a_1 a^\dag_2)\ ,
\end{equation}
where $a_{1,2}$, $a_{1,2}^\dag$ annihilate and create atoms in the first, respectively
second well and satisfy the Bose commutation relations $[a_i\,,\,a^\dag_j]=\delta_{ij}$.
The three contributions in the above hamiltonian correspond to
a hopping term dependent on the tunneling
amplitude $T$, and on two on-site energy terms, 
one quadratic in the number operator with coupling constant $U$,
describing the boson-boson repulsive contact interaction, 
the other due to the trapping potential, with the parameter $E_1$, $E_2$ representing the wells depth.

Notice that the total number $N$ of particles is conserved by the Hamiltonian (\ref{BH}).
As a consequence, the Hilbert space of the system is $N+1$-dimensional and can be 
spanned by the set of Fock states, describing the situation in which the first well is filled with
$k$ particles, while the other contains $N-k$ atoms; they are generated by the action of 
the creation operators on the vacuum:
\begin{equation}
\label{Fock}
|N; k\rangle = \frac{(a_1^\dagger)^k (a_2^\dagger)^{N-k}}{\sqrt{k! (N-k)!}}\ |0\rangle\ ,
\qquad k=0, 1,\ldots,N\ .
\end{equation}

Alternatively, one can introduce coherent-like states, depending on two
real parameters, an amplitude $\xi\in[0,\ 1]$ and a phase $\varphi\in[0,\ 2\pi]$,
\begin{equation}
\label{coherent}
|N; \xi,\varphi\rangle = \frac{1}{\sqrt{N!}}\left(\sqrt\xi\, e^{i \varphi/2}a^\dag_1\,+\,
\sqrt{1-\xi}\, e^{-i \varphi/2} a^\dag_2\right)^N\ |0\rangle\ .
\end{equation}
They describe situations in which all $N$ particles are in a coherent superposition,
with definite relative phase $\varphi$ and relative mean occupation number 
$\langle N; \xi, \varphi |\big(a^\dag_1 a_1- a^\dag_2 a_2\big) |N; \xi,\varphi\rangle=
1-2\xi$. The set of states (\ref{coherent}) is overcomplete,
\begin{equation}
\label{completeness}
\int_0^1 d\xi \int_0^{2\pi}\frac{d\varphi}{2\pi}\ |N; \xi,\varphi\rangle \langle N; \xi, \varphi |
=\frac{1}{(N+1)}\mathbbm 1_{N+1}\ ,
\end{equation}
and although not strictly orthogonal, they become so in the limit of large $N$:
\begin{equation}
\label{orthogonality}
\langle N; \xi, \varphi |N; \xi',\varphi'\rangle \approx \frac{1}{N}\delta(\xi-\xi')\,
\delta(\varphi-\varphi')\ .
\end{equation}
Nevertheless, one can always express any Fock state (\ref{Fock}) 
in terms of coherent states by using the completeness relation (\ref{completeness}),
\begin{equation}
\label{Fock2}
|N;k\rangle=(N+1)\, \int_0^1 d\xi \int_0^{2\pi}\frac{d\varphi}{2\pi}\ 
\langle N; \xi, \varphi | N; k\rangle\ |N; \xi,\varphi\rangle\ ,
\end{equation}
where the overlap functions are explicitly given by
\begin{equation}
\label{overlap} 
\langle N; \xi, \varphi | N; k\rangle=
\binom{N}{k}^{1/2}\, \xi^{\frac{k}{2}}\, (1-\xi)^{\frac{N-k}{2}}\,
e^{-i\varphi\left(k-\frac{N}{2}\right)}\ .
\end{equation}

As well known \cite{Lewenstein}, the Hamiltonian (\ref{BH}) describes a cross-over
between a superfluid and insulator phases, which becomes a true quantum phase transition,
with order parameter depending on the ratio $T/U$, in the limit of an infinite number of wells. 
Neglecting the shift $E=E_1-E_2$,
when an atom hops between the two wells, 
it loses an energy $T$ and gains instead an energy $U$; thus, when $T/U\ll 1$, 
this situation is energetically suppressed and, at equilibrium, the ground state of the system is given by
a Fock state with equal number $N/2$ of bosons per well, $|N; N/2\rangle$ (Mott insulator phase).
On the other hand, when $T/U\gg 1$, the system shows phase coherence; all the $N$ particles are in
the same superposition and the ground state of (\ref{BH}) can be approximated by
$|N; \xi,\varphi\rangle$, with a definite relative phase and occupation number (superfluid phase).

\section{Generalized quantum measure}

As explained in the introductory remarks, in a typical experimental set up the double well trap
is filled with a large number $N$ of cold atoms, whose dynamics is driven by the Hamiltonian (\ref{BH}).
Since in-trap measures of relevant system observables are difficult, one can obtain indirect
information on the atom behaviour by switching off the trapping potential and letting the atoms 
expand freely. The two fractions of atoms coming from the two wells will eventually overlap;
one is then able to observe the spatial distribution of the atoms by illuminating the
overlapping clouds with a probe light and by registering the corresponding absorption image.

In order to model this process of measure, we shall use a second-quantized many-body formalism.
Let us first introduce a complete set of single-particle atom states $\{|W_i\rangle\}_{i=1}^\infty$,
obtained by acting with the set of all creation operators $a_i^\dagger$ on the vacuum,
$|W_i\rangle\equiv a_i^\dagger\, |0\rangle$. Although just two creation operator 
$a_1^\dagger$ and $a_2^\dagger$ are enough to
properly describe an atom confined into the two wells, in absence of the trapping potential
a single atom moves freely in space and therefore a complete set of wave functions
$W_i(x)\equiv\langle x|W_i\rangle = \langle x | a^\dagger_i |0\rangle$
is needed for properly representing its states ($x$ represents the collection of spatial coordinate variables). The creation operator $\psi^\dagger(x)$ of an atom at
position $x$, $\psi^\dagger(x)\, |0\rangle=|x\rangle$, can then be decomposed as
\begin{equation}
\label{psi}
\psi^\dagger(x)=\sum_{i=1}^\infty \overline{W}_i(x)\, a^\dagger_i\ .
\end{equation}
Since the single particle states are orthonormal, one can invert this relation and equivalently write
\begin{equation}
\label{a}
a^\dagger_i=\int dx\, \overline{W}_i(x)\, \psi^\dagger(x)\ .
\end{equation}
Further, from $[a_i,\, a^\dagger_j]=\langle W_i | W_j\rangle=\delta_{ij}$, one recovers the
standard bosonic (equal-time) commutation relations:
\begin{equation}
\big[\psi(x),\ \psi^\dagger(x')\big]=\delta(x-x')\ .
\end{equation}
With this formalism, one can now easily compute the action of the operator $\psi(x)$
on a many-body state $|N;\xi,\varphi\rangle$. Using the above commutation relations,
one easily finds:
\begin{equation}
\label{action}
\psi(x)|N;\xi,\varphi\rangle={\sqrt N}\, \left(\sqrt\xi\, W_1(x)e^{i\varphi/2}\,+\,
\sqrt{1-\xi}\, W_2(x)\, e^{-i\varphi/2}\right)|N-1;\xi,\varphi\rangle\ ,
\end{equation}
so that the wave function associated to the coherent state $|N;\xi,\varphi\rangle$, depending
on the positions $x_1,\ x_2,\ldots,\ x_N$ of the $N$ atoms, is simply given by
\begin{equation}
\label{ }
\langle x_1, x_2, \ldots, x_N|N;\xi,\varphi\rangle=\sqrt{N!}\,\prod_{i=1}^N
\left(\sqrt\xi\, W_1(x_i)e^{i\varphi/2}\,+\,
\sqrt{1-\xi}\, W_2(x_i)\, e^{-i\varphi/2}\right)\ .
\end{equation}
Similarly, the action of the destruction operator $\psi(x)$ for an atom at position $x$ 
on a Fock state $|N; k\rangle$
can be obtained from (\ref{action}) through the expansion (\ref{Fock2}).

Strictly speaking, all this holds at fixed time, {\it e.g.} at $t=\,0$. 
However, since the atoms are assumed to evolve freely
once the trap is released, insertion of the time dependence in the formalism is straightforward.
Assume that the confining double well potential is released at time $t=\,0$ and denote by
$U_t$ the unitary operator that evolves freely in time the initial one particle states:
\begin{equation}
\label{evolution}
|W_i(t)\rangle\equiv U_t |W_i\rangle = U_t\, a^\dagger_i |0\rangle
= a^\dagger_i(t) |0\rangle\ ,\qquad a^\dagger_i(t)=U_t\, a^\dagger_i\, U_t^\dagger\ .
\end{equation}
The corresponding wave function is given by
$W_i(x;t)\equiv\langle x|W_i(t)\rangle = \langle x | a^\dagger_i(t)|0\rangle$, and coincides
with a transformed $W_i(x)$ under a ballistic expansion.
Since the dynamics is free, every particle in a many-body state will evolve independently with $U_t$;
therefore, the evolution up to time $t$ of the $t=\,0$ coherent state $|N;\xi,\varphi\rangle$
will simply be given by
\begin{equation}
\label{coherent-t}
|N; \xi,\varphi; t\rangle = \frac{1}{\sqrt{N!}}\left(\sqrt\xi\, e^{i \varphi/2}a^\dag_1(t)\,+\,
\sqrt{1-\xi}\, e^{-i \varphi/2} a^\dag_2(t)\right)^N\ |0\rangle\ ,
\end{equation}
and similarly for a Fock state (\ref{Fock}).

In this picture, states evolve in time while observables remain fixed and thus the operator $\psi(x)$
results time-independent. Since at each instant of time $t$ the collection $\{|W_i(t)\rangle\}_{i=1}^\infty$ 
is a complete set of single particle states obtained from the vacuum by the action of the creation operators
$a^\dagger_i(t)$, $\psi(x)$ can be equivalently decomposed as
$\psi(x)=\sum_{i=1}^\infty {W}_i(x;t)\, a_i(t)$ for all times.
The action of $\psi(x)$ on $|N; \xi,\varphi; t\rangle$ is then as in
(\ref{action}), once expressed in terms of the evolved wave functions:
\begin{equation}
\label{action-t}
\psi(x)|N;\xi,\varphi;t\rangle={\sqrt N}\, \left(\sqrt\xi\, W_1(x;t)e^{i\varphi/2}\,+\,
\sqrt{1-\xi}\, W_2(x;t)\, e^{-i\varphi/2}\right)|N-1;\xi,\varphi;t\rangle\ .
\end{equation}
Through this result, one can easily compute the mean value of the density operator at point $x$,
\begin{equation}
\label{density}
n(x)\equiv \psi^\dagger(x)\,\psi(x)\ ,
\end{equation}
in any many-body state.
For instance, if the atoms just before the release of the confining potential
are in a superfluid state described by a state
$|N;\xi,\varphi\rangle$, after a time $t$ of free evolution of the cloud one
would obtain:
\begin{eqnarray}
\label{average-density}
\nonumber
\langle n(x)\rangle_{\xi,\varphi}=\langle N;\xi,\varphi;t | n(x)|N;\xi,\varphi;t\rangle &=&
N\Big[ \xi\, |W_1(x;t)|^2 + (1-\xi)\, |W_2(x;t)|^2 \\
 &+&2\sqrt{\xi(1-\xi)}{\cal R}e\Big( W_1(x;t) \overline{W}_2(x;t)\, e^{i\varphi}\Big)\Big]\ ,
\end{eqnarray}
exhibiting the well known interference pattern.
More in general, if at $t=\,0$, the time at which the confining potential is switched off,
the system is in a generic many body state described by the density matrix $\rho$,
the average density after a free evolution of the system up to time $t$,
$\rho\mapsto \rho(t)$, will be given by
\begin{equation}
\label{trace}
\langle n(x)\rangle_{\rho(t)}={\rm Tr}\big[ n(x)\, \rho(t) \big]\ .
\end{equation}

As always in quantum mechanics, this mean value
refers to statistical averages over many experimental runs, where each time the system is prepared
at $t=\,0$ in the same state $\rho$. 
However, as explained before, this is not
what it is done in a typical experiment, since the direct measure of the observable $n(x)$ is
usually problematic. Instead, one takes a picture of the released atom cloud at time $t$, 
collecting the resulting absorption image,
while loosing the sample. 

When the two clouds can be considered independent or described by a state like
$|N/2;N/2\rangle$, this procedure is commonly assumed to correspond to a
projection onto a fixed-phase state, with definite $\xi=1/2$
\cite{Leggett,Stringari}.
In the following, we extend this treatment by considering projections of the form
$|N;\xi,\varphi;t\rangle \langle N;\xi,\varphi;t |$ onto a state
with a definite relative phase $\varphi$ and amplitude $\xi$, although randomly given,
producing the average density $\langle n(x)\rangle_{\xi,\varphi}$%
\footnote{This result can be naively understood by interpreting the formation of the absorption image
as the result of the interaction of the system with a classical, macroscopic measuring
apparatus: many atoms concur to the formation of a single pixel in the image and this is possible
only if all atoms are in a same coherent superposition; as explained before, this situation is described 
by the coherent states (\ref{coherent}), which in turn are the more classical
among all quantum states.}
\!: our purpose is to compute the average over all the absorption images within this approach
and to compare it with the expression (\ref{trace}).
As it will become clear below, the inclusion also of the variable $\xi$
is necessary to preserve the overall probability.
Further, if the whole experiment is repeated, preparing the system in the same state $\rho$ and taking
its picture after a free expansion up to the same time $t$, a different phase and amplitude
will be selected. The distribution of the obtained values over many repetitions
is determined by the initial state $\rho$ through the probability
$ \langle N;\xi,\varphi;t | \rho(t) |N;\xi,\varphi;t\rangle$, which indeed gives the weight
in $\rho$ of the state configuration with fixed phase $\varphi$ and amplitude $\xi$.

For large number of atoms $N$,
this description is perfectly in agreement with what it is experimentally measured through the
absorption images, that exhibit interference patterns irrespectively
from the initial state, being either superfluid or Mott.
Indeed, in many-body physics, one can assimilate ensemble averages with
mean values with respect to macroscopically occupied many-body states, provided the number of particles involved
is large enough \cite{Feynman}. Therefore, the larger the number $N$ of atoms the system contains, 
the better a {\it single} absorption image will model the average of the density
operator $n(x)$ in the state $|N;\xi,\varphi;t\rangle$, thus
reproducing the interference pattern $\langle n(x)\rangle_{\xi,\varphi}$
given in (\ref{average-density}) \cite{Demler}.

In a typical experiment, all obtained absorption images are subsequently superimposed,
giving rise to the average
\begin{equation}
\label{mean1}
\langle n(x)\rangle=\frac{1}{\cal N}\sum_{i=1}^{\cal N}\ 
\langle n(x)\rangle_{\xi_i,\varphi_i}\ ,
\end{equation}
with $\cal N$ the total number of taken pictures. In this sum, any
specific density pattern $\langle n(x)\rangle_{\xi_i,\varphi_i}$, with
values $\xi_i$ and $\varphi_i$ of amplitude and phase, will occur $\nu_i$ times,
{\it i.e.} with a frequency $p_i\equiv\nu_i/{\cal N}$; the average (\ref{mean1}) may then be more
conveniently rewritten as
\begin{equation}
\label{mean2}
\langle n(x)\rangle=\sum_\alpha\ p_\alpha\ 
\langle n(x)\rangle_{\xi_\alpha,\varphi_\alpha}\ ,
\end{equation}
where the sum is now over the set of {\it distinct} absorption images.
In the limit of large $\cal N$, the above sum becomes an integral over all possible values of
amplitude and phase. Therefore, the procedure of averaging over all obtained images really
corresponds to summing the mean values of $n(x)$
over all states $|N;\xi,\varphi;t\rangle$ with definite $\xi$ and $\varphi$, 
each contribution weighted with the associated occurrence probability
$ p_\alpha\sim\langle N;\xi,\varphi;t | \rho(t) |N;\xi,\varphi;t\rangle$, determined by the initial state $\rho$.
Mathematically, this procedure is then described by the operation of trace of $n(x)$
over the density matrix
\begin{equation}
\label{V-rho}
\tilde\rho(t)=
\frac{N+1}{2\pi}\int_0^1 d\xi \int_0^{2\pi}d\varphi\ 
\langle N;\xi,\varphi;t | \rho(t) |N;\xi,\varphi;t\rangle\
|N; \xi,\varphi;t\rangle \langle N; \xi, \varphi;t |\ .
\end{equation}
More precisely, the transformation $\rho\mapsto \tilde\rho$ defines
a linear map ${\cal V}_N$ acting on the space of all density matrices: it is a realization
of a Positive Operator Valued Measure (POVM) \cite{Takesaki,Peres,Alicki}, a generalization of the
more standard von Neumann measure:%
\footnote{Strictly speaking, after the release of the trapping potential, the integral 
${\cal P}_{N+1}\equiv\sqrt{\frac{N+1}{2\pi}}\int_0^1 d\xi \int_0^{2\pi}d\varphi\, V(N;\xi,\varphi;t)$ 
gives the projector operator on the $(N+1)$-dimensional subspace spanned by the vectors
(\ref{Fock}) and not the identity $\mathbbm 1$ over the whole Hilbert space; therefore,
the POVM should be more correctly defined by the set
$\{ V(N;\xi,\varphi;t),\ {\mathbbm 1}-{\cal P}_{N+1}\}$.
However, since all system states at time $t$ come from the free evolution of states 
belonging to this subspace, the action of ${\mathbbm 1}-{\cal P}_{N+1}$ has no effect, and the definition
(\ref{V-rho2}) given below follows.}
\begin{equation}
\label{V-rho2}
\tilde\rho(t)\equiv{\cal V}_N(\rho(t))=
\int_0^1 d\xi \int_0^{2\pi}d\varphi\ V(N;\xi,\varphi;t)\ \rho(t)\
V(N;\xi,\varphi;t)\ ,
\end{equation}
with
\begin{equation}
V(N;\xi,\varphi;t)=
\sqrt{\frac{N+1}{2\pi}}\ 
|N; \xi,\varphi;t\rangle \langle N; \xi, \varphi;t |\ .
\end{equation}
The proper quantum mechanical expression for the matter density profile resulting 
from the average over all absorption images
is then given by: 
\begin{equation}
\label{mean-density}
\langle n(x)\rangle=
\frac{N+1}{2\pi}\int_0^1 d\xi \int_0^{2\pi} d\varphi\ 
\langle N;\xi,\varphi;t | \rho(t) |N;\xi,\varphi;t\rangle\
\langle N;\xi,\varphi;t | n(x) |N;\xi,\varphi;t\rangle\
\ ,
\end{equation}
which is clearly different from the average $\langle n(x)\rangle_{\rho(t)}$ 
in (\ref{trace}), that is usually adopted in interpreting the experimental data
\cite{Stringari}-\cite{Leggett-book}, \cite{Leggett}.

The difference between these two expressions
of the mean matter density may have experimental relevance.
In this respect, the most interesting case is represented by the Fock states (\ref{Fock}),
describing situations for which the system has a definite number of atoms 
in the two wells: they can be easily realized in practice
by raising the barrier between the two traps. Let us assume that, 
just before the release of the trapping potential, the system be prepared
in a state containing $k$ atoms in the first well and $N-k$ in the second, so that
\begin{equation}
\label{rho-Fock}
\rho=|N;k\rangle\langle N;k|\ .
\end{equation}
In the limit of large $N$, the distribution of matter
in a single absorption image taken after a free evolution up to time $t$ is still given by
(\ref{average-density}), with a randomly picked relative phase $\varphi$ and amplitude $\xi$.
Nevertheless, while the distribution of the possible values of $\varphi$
over many images remains flat, that of $\xi$ follows
a time-independent binomial law:
\begin{equation}
\label{binomial}
\langle N;\xi,\varphi;t | \rho(t) |N;\xi,\varphi;t\rangle=
\Big|\langle N;\xi,\varphi;t |N;k;t\rangle\Big|^2=
\binom{N}{k}\, \xi^k\, (1-\xi)^{N-k}\ .
\end{equation}
Therefore, if the experiment is repeated many times, each single absorption image will show
interference fringes with the same spatial period but with randomly distributed offset position,
while the amplitude $\xi$ remains essentially constant through all images,
since for large $N$ the distribution (\ref{binomial})
is peaked around its average value. As a consequence, by superimposing all absorption pictures, the interference pattern is averaged away, as it is easily found by inserting (\ref{rho-Fock}) into
the expression (\ref{mean-density}); indeed, by performing the integrals, one explicitly finds:
\begin{equation}
\label{final1}
\langle n(x)\rangle
=\frac{N}{N+2}\Bigl[(k+1)\,|W_1(x,t)|^2\,+\,(N-k+1)\,|W_2(x,t)|^2\Bigr]\ .
\end{equation}
On the other hand, by adopting the definition (\ref{trace}), using (\ref{rho-Fock}), one would instead obtain 
\begin{equation}
\label{final2}
\langle n(x)\rangle_{k}
=k\,|W_1(x,t)|^2\,+\,(N-k)\,|W_2(x,t)|^2\ .
\end{equation}
This two expressions coincide only for equally filled wells, $k=N/2$; since this is
the situation encountered in most experimental setups, this explains why 
in all discussions the definition (\ref{trace}) for the average density has been
adopted instead of the quantum mechanically correct one given in (\ref{mean-density}).

Further, notice that the difference between the above two expressions
is not vanishing as $N$ becomes large,
\begin{equation}
\label{difference}
\langle n(x)\rangle-\langle n(x)\rangle_{k}\simeq
\left(|W_1(x,t)|^2\,-\,|W_2(x,t)|^2\right)+ O(1/N)\ ,
\end{equation}
and therefore it is of relevance in certain experimental situations. This is the case
for states with very unbalanced fillings of the two wells, for which $k$ is very small.
By selecting a time of flight $t$ for which the atom cloud is sufficiently spread out
to be visible in the absorption images, but such that
the two wave functions $W_1(x,t)$ and $W_2(x,t)$
are still sufficiently well separated in space, 
one should be able to experimentally measure the weight in front of the two
contributions $|W_1(x,t)|^2$ and $|W_2(x,t)|^2$ and thus 
quantitatively test the validity of the assumption of the projection 
on coherent states during the absorption imaging process. 

As a further remark, notice that if instead the system is prepared 
at $t=\,0$ in a superfluid state,
described by the density matrix
\begin{equation}
\label{rho-superfluid}
\rho=|N;\xi',\varphi'\rangle\langle N;\xi',\varphi'|\ ,
\end{equation}
all absorption images taken after a time $t$ will show exactly the same
interference pattern, described by the given amplitude $\xi'$ and phase $\varphi'$.
This is a consequence of the (large $N$) orthogonality of the coherent states
$|N;\xi,\varphi,\rangle$ as given by (\ref{orthogonality}), which remains
true for all $t$, since the time evolution is unitary. By superimposing
all taken images, using (\ref{mean-density}) and (\ref{orthogonality})
one then finds
\begin{equation}
\langle n(x)\rangle=
\langle N;\xi',\varphi';t | n(x) |N;\xi',\varphi';t\rangle\ ,
\end{equation}
so that the obtained average density coincides with the mean operator density,
thus reproducing the interference pattern given in (\ref{average-density}),
as observed in actual experiments.

Finally, let us notice that if the double-well trap is filled with non-interacting
fermions instead of bosons, no inteference pattern can emerge in a single absorption
image, since the fermions can not macroscopically occupy the same state.
However, if a small attractive interaction is switched on, such that the new
ground state is BCS-like with a sufficient number of Cooper-pairs, then the description
of the system in terms of the Bose-Hubbard
Hamiltonian (\ref{BH}) holds also in this case \cite{Makhlin,Fazio,Bruder}, 
and, as a consequence, it is expected that single
absorption images will show inteference fringes.

\vskip 1cm

\noindent
{\bf Acknowledgements}

\noindent
Discussions with F. Piazza and A. Smerzi and G. Modugno, 
C. Fort and L. Fallani of the Quantum Degenerate Gases group at LENS are gratefully 
acknowledged. S.A. acknowledges Eurotech s.p.a. for finantial support.
This work is supported by the MIUR project 
``Quantum Noise in Mesoscopic Systems''.


\end{document}